\RequirePackage{fix-cm}
\documentclass[smallextended]{svjour3}       
\smartqed  
\usepackage{graphicx}
\usepackage{color}
\usepackage{amsmath}
\usepackage{mathrsfs}
\usepackage{psfrag}
\begin{document}

\title{
Iterative control of G\"{o}rtler vortices via local wall deformations 
}


\author{Adrian Sescu         \and
           Lamiae Taoudi         \and
           Mohammed Afsar 
}


\institute{F. Author \at
              Department of Aerospace Engineering, Mississippi State University, 330 Walker at Hardy Rd, Mississippi State,  MS 39762 \\
              Tel.: +662-325-7484\\
              Fax: +662-325-3623\\
              \email{sescu@ae.msstate.edu}           
           \and
           S. Author \at
              Department of Aerospace Engineering, Mississippi State University, 330 Walker at Hardy Rd, Mississippi State,  MS 39762
           \and
           T. Author \at
              Mechanical \& Aerospace Engineering Department, University of Strathclyde, 75 Montrose St, Glasgow G1 1XQ, UK
}

\date{Received: date / Accepted: date}

\maketitle

\begin{abstract}
G\"{o}rtler vortices develop along concave walls as a result of the imbalance between the centrifugal force and radial pressure gradient. In this study, we introduce a simple control strategy aimed at reducing the growth rate of G\"{o}rtler vortices by locally modifying the surface geometry in spanwise and streamwise directions. Such wall deformations are accounted in the boundary region equations (BRE) by using a Prandtl transform of dependent and independent variables. The vortex energy is then controlled via a classical proportional control algorithm for which either the wall-normal velocity or the wall shear stress serves as the control variable. Our numerical results indicate that the control algorithm is quite effective in minimizing the wall shear stress.
\end{abstract}

\section{Introduction}

The control of disturbances in laminar boundary layers (or fully-developed turbulent boundary layers) targets a reduction in the energy carried the streamwise oriented structures that feature high- and low-velocity streaks in the near wall region. These streaks are known to be the starting points of the so-called `bursting sequence' prior to transition from laminar to turbulent flow. One popular control strategy involves using regions of localized suction below low-velocity streaks and blowing regions of high-velocity streaks. The net result of this approach is a decrease in the spanwise variation of the streamwise velocity and, therefore, a commensurate reduction in disturbance energy or wall shear stress.  

In this paper, we study the control of boundary layer streaks in the form of Gortler instabilities that exist owing to an imbalance between centrifugal forces and radial pressure gradients inside a boundary layer developing on a concave surface. As a result of the imbalance between centrifugal forces and radial pressure gradients, low-speed fluid particles from the wall are lifted up while high-speed particles from above are moving toward the wall, generating elongated streamwise streaks inside the boundary layer flow (\cite{Gortler}, \cite{hall1}, \cite{hall2}, \cite{saric}). The upwelling and downwelling of fluid particles as they move downstream at different velocities suggests that it is possible to control the amplification of the vortices by using wall effects, such as transpiration (as considered by previous authors) or local surface deformation that we consider here. For highly curved walls, vortex formation occurs rapidly (with respect to the freestream flow timescale) and can significantly alter the mean flow causing the laminar flow to break down into turbulence. G\"{o}rtler \cite{Gortler}, in his seminal paper, assumed that the flow was locally parallel and used a normal mode analysis to prove that vortex formation occurs when the dimensionless parameter, $Re \sqrt{\delta^* / R_0^*}$ (referred to as the G\"{o}rtler number), exceeds a critical value ($Re$ is Reynolds number in terms of the free-stream velocity and the boundary layer thickness, $\delta^*$, and $R_0^*$ is the radius of curvature). Later, Hall \cite{hall1,hall2}, found that the stability of G\"{o}rtler vortices depends on the amplitude, location and form of the upstream disturbance indicating that there is no unique neutral stability curve for this problem. The initial value problem formalism is adopted throughout this paper.

The literature about boundary layer control is rich and we do not intend to mention all previous studies that were performed. It is important to note, however, that a closely related study is the active wall control that was applied in the context of turbulent channel flow by Choi et al. \cite{Choi} as a means to reduce skin friction drag. They performed direct numerical simulations (DNS) with active wall control based on blowing and suction informed by indicators placed inside the flow in a detection plane close to the wall (approximately 10 wall units above the wall), and obtained a drag reduction of approximately 25\%. Koumoutsakos \cite{Koumoutsakos1,Koumoutsakos2}, similarly, introduced a feedback control algorithm using only wall information in simulations of turbulent channel flow at low Reynolds number. By using the vorticity flux components as inputs to the control algorithm (which can be obtained as a function of time) and by measuring the instantaneous pressure at the wall and calculating its gradient, a skin friction reduction of 40\% was claimed. Optimal control of boundary layer disturbances based on wall blowing and suction has been also applied by several other research groups (see, for example, Joslin et al. \cite{Joslin}, Bewley and Liu \cite{Bewley}, Lee et al. \cite{Lee}, Walther et al. \cite{Walther}, Corbett and Bottaro \cite{Corbett}, Hogberg and Henningson \cite{Hogberg}, Zuccher et al. \cite{Zuccher}, Cherubini et al. \cite{Cherubini}, Papadakis et al. \cite{Papadakis}, to name a few).

Controlled wall deformations counteracting streaks in turbulent boundary layers and consequently reducing the wall skin friction were considered in several previous studies. Endo et al. \cite{Endo}, for example, performed DNS studies with feedback control of deformable walls to reduce the skin friction in a turbulent channel flow. The control scheme was based on physical arguments pertaining to the near-wall coherent structures, and provided a 10\% friction drag reduction. Endo et al. \cite{Endo} also pointed out that the energy input required to deform the wall is much smaller than the pumping power required for suction/blowing. Kang et al. \cite{Kang} investigated the potential of reducing the skin-friction drag in a turbulent channel flow via active wall motions. Interestingly, they noticed that the instantaneous wall surface shape also took the form of elongated streaks as in laminar boundary layers. A reduction of the friction drag on the order of 13-17\% was realized by their approach. Koberg \cite{Koberg} experimentally investigated an approach for reducing skin friction in a turbulent flow via active wall deformation. They attempted to match the velocity sensed away from the wall by imposing a velocity of opposite direction at the wall; a skin friction reduction of 15\% was accomplished here.

\begin{figure}
 \begin{center}
     \includegraphics[width=8cm]{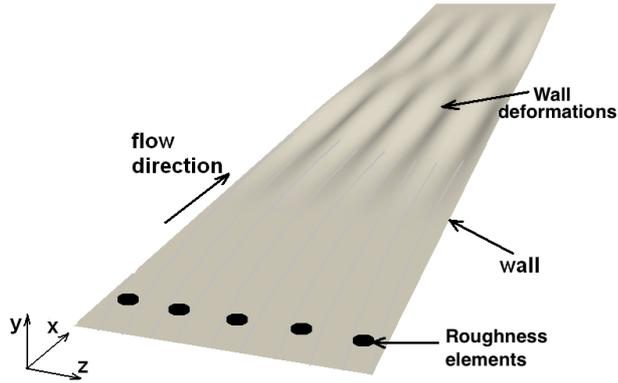}
 \end{center}
  \caption{\label{} Sketch of the boundary layer and the deformed wall}
  \label{surf}
\end{figure}

In the present study, we show that a simple flow control strategy based on wall deformation can significantly reduce the energy of G\"{o}rtler vortices developing within a laminar boundary layer along a concave surface. The type of control falls in the category of `opposition control', as introduced by Choi et al.; however, rather than wall transpiration we use local wall deformation (as in turbulent boundary layer studies of Kang et al. \cite{Kang} and Endo et al. \cite{Endo}). It is shown that the control can be implemented self-consistently using the high Reynolds number asymptotic framework for a laminar boundary layer flow on a curved wall,
in which the flow is governed by the so-called boundary region equations. Local changes in the surface geometry can be introduced into these equations conveniently through a Prandtl transform (Yao \cite{Yao}) that does not alter the parabolic character of the basic equations themselves, and therefore allows solution by a numerical marching technique. The variations in local surface shape then allows for a relatively straightforward control strategy to be implemented to the transformed set of equations in order to determine the optimum wall deformation that reduces the vortex energy. The streaks are initiated by perturbing the upstream flow with a periodic array of roughness elements that enters the analysis through initial conditions that we obtain from a previously derived asymptotic solution by Goldstein et al. \cite{Goldstein1,Goldstein2} (the actual flow around roughness elements is not needed to be modeled in this work). The wall deformation is then applied using a control algorithm based on a proportional controller, with the control variable being either the wall-normal velocity disturbance in a plane section that is parallel to the wall or the wall shear stress. Figure \ref{surf} shows a sketch of the boundary layer and the deformed wall in the downstream region.

\section{Boundary region equations}\label{}

To fix ideas, we consider an incompressible boundary layer flow over a concave surface, with upstream perturbations provided by a spanwise periodic array of roughness elements at some streamwise location, $x = x_0$ from the leading edge. The spanwise length scale of the roughness row, $\Lambda$, is in the same order of magnitude as the local boundary-layer thickness $\delta \equiv x_0/ \sqrt{R} = x_0 \delta$ at the roughness location $x_0$, where $R$ is the Reynolds number based on roughness location the free stream velocity, and $\delta \equiv R^{-1/2}$ being the boundary layer thickness scaled by the roughness spanwise separation, which is an $O(1)$ length scale. As mentioned earlier, the detailed flow around the roughness elements is not modeled here, since only the initial/upstream conditions are required for the input to boundary region equations. The upstream conditions are provided by an asymptotic solution as derived in Goldstein et al. \cite{Goldstein1}, where the height of the roughness element was assumed small relative to the spanwise separation.

We adopt a body-fitted coordinate system, where the original Navier Stokes equations are transformed according to Lam\'{e} coefficients, $h_{1}=(R_{0}-y)/R_{0},h_{2}=1$, and the radius of curvature, $R_0$, is much larger than the spanwise separation of the roughness elements; i.e. $R_0 \gg O(1)$. The origin of the coordinate system is located at the leading
edge, with the streamwise $x$-axis aligned with the wall surface,
$y$-axis perpendicular to the wall surface, and $z$-axis aligned with the
spanwise direction. The velocity field normalized by the freestream velocity, and the pressure normalized by the freestream dynamic pressure are expand as
\begin{equation}\label{zzz}
\{ \tilde{u}, \tilde{v}, \tilde{w}, \tilde{p} \} = \{ u(X,y,z), \varepsilon v(X,y,z), \varepsilon w(X,y,z), \varepsilon^2 p(X,y,z) \} + ...
\end{equation}
where $\varepsilon = 1/R_{\Lambda}$, $X = x/R_{\Lambda}$ is the slow streamwise variable, and $(y,z)$ are $O(1)$ wall-normal and spanwise coordinates, respectively. 
As in Wu et al. \cite{Wu}, G\"{o}rtler vortices are expected to
develop at $x \sim  \Lambda^{*} R_{\Lambda}$ (where $ \Lambda^{*}$ is roughness spanwise separation), and are governed by the nonlinear boundary region equations (BRE)
\begin{equation}\label{nq1}
u_{X}+v_{y}+w_{z}=0,
\end{equation}
\begin{equation}\label{nq2}
uu_{X}+vu_{y}+wu_{z}=u_{yy}+u_{zz},
\end{equation}
\begin{equation}\label{nq3}
uv_{X}+vv_{y}+wv_{z}+G_{\Lambda}u^{2}=-p_{y}+v_{yy}+v_{zz},
\end{equation}
\begin{equation}\label{nq4}
uw_{X}+vw_{y}+ww_{z}=-p_{z}+w_{yy}+w_{zz},
\end{equation}
where the effect of the wall curvature is contained in the term involving the global G\"{o}rtler number $G_{\Lambda}=R_{\Lambda}^{2}/R_{0}$ in equation (\ref{nq3}). The absence of streamwise second order derivatives in the BRE indicates that they are parabolic in the streamwise direction and can be solved numerically using a space-marching technique (Hall \cite{hall1,hall2}). Goldstein et al. \cite{Goldstein1} showed that if the roughness height is small relative to the roughness spanwise separation the downstream flow can be solved using the linearized boundary region equations (which are also parabolic in the streamwise direction). Therefore, we use the upstream conditions derived in Goldstein et al. \cite{Goldstein1} as the input to the BRE system (\ref{nq1}-\ref{nq4}).

Prandtl transform (or Prandtl transposition theorem, Yao \cite{Yao}) is conveniently applied to the BRE to incorporate local changes in wall surface geometry defined through the function $\mathscr{F}(X,y)$. The new wall normal variable and perturbation velocity are defined by $
Y = y - \mathscr{F}
$ and $
\hat{v} = v - \left( u \mathscr{F}_X + w \mathscr{F}_z \right)
$, respectively. Using the chain-rule, the transformed BREÕs are therefore easily obtained as
\begin{equation}\label{neq1}
u_{X}+\hat{v}_{Y}+w_{z}=0,
\end{equation}
\begin{equation}\label{neq2}
uu_{X}+\hat{v}u_{Y}+wu_{z}=u_{YY}+\left( \partial_z - \mathscr{F}_z \partial_Y \right)^2 u - \mathscr{F}_{zz} u_Y,
\end{equation}
\begin{equation}\label{neq3}
uv_{X}+\hat{v} v_{Y}+w v_{z}+G_{\Lambda}u^{2}=-p_{Y}+v_{YY}+\left( \partial_z - \mathscr{F}_z \partial_Y \right)^2 v - \mathscr{F}_{zz} v_Y,
\end{equation}
\begin{equation}\label{neq4}
uw_{X}+\hat{v}w_{Y}+ww_{z}=-\left( \partial_z - \mathscr{F}_z \partial_Y \right) p+w_{YY}+\left( \partial_z - \mathscr{F}_z \partial_Y \right)^2 w - \mathscr{F}_{zz} w_Y,
\end{equation}
The transformed BRE obviously retains its parabolic character and can, therefore, be solved by the same marching technique (see Goldstein et al. \cite{Goldstein1} or Sescu and Thompson \cite{Sescu1}). In this paper, $\mathscr{F}(X,z)$ is a continuous and smooth function that will be obtained at discrete points $(X,z)$ by iterations within the control algorithm, where $\mathscr{F}(X,z)=0$ corresponds to the original undeformed surface.

The boundary conditions at the wall are given as
$
u(X,y_{r},z)=v(X,y_{r},z)=w(X,y_{r},z)=0,
$
where the surface of the wall is described by the function $y_{r}= \mathscr{F}(X,z)$. At the top of the domain, vanishing gradients are imposed on all dependent variables. The surface deformations are initiated at a specified streamwise location on the surface using a ramping function of the form
$
0.5[1-\cos(\pi(X-X_1)/(X_2-X_1))]
$ applied between $X_1$ and $X_2$. The surface gradients in the streamwise direction are very small because a smooth ramping function is applied to initiate the surface deformations over a long streamwise distance; note that the ratio between the roughness elements spanwise separation and the ramping distance is in the order of $0.01$, which suggests that the streamwise pressure gradient and the streamwise second order derivative of $u$ are small compared to spanwise or wall-normal second order derivatives over the length which the deformation is applied (this means that $\partial p/ \partial x$ and $\partial^2 u/ \partial x^2$ are absent from the first momentum equation). This is likely to be true as long as the streamwise length scale of the deformation is much larger than the triple deck scale, which indicates that the BRE will continue to remain parabolic since the streamwise flow will depend on the slow streamwise coordinate and the flow will expand as equation (\ref{zzz}).

\section{Proportional control algorithm}

A simple control algorithm is utilized here to determine the functional form $\mathscr{F}(X,z)$, through an iterative loop using control variables from the transformed BREs. The aim is to vary $\mathscr{F}(X,z)$ such that the energy associated with the G\"{o}rtler vortices is minimized. The control variable is either the wall-normal velocity disturbance in a $y = const$ plane at a specified height above the wall surface, or the wall shear stress.
A typical proportional controller is considered here as 
$
\mathscr{A}(X,z) = K_p*e(X,z)$, where $K_p$ is the proportional gain, and $
e(X,z) = v(X,y_{c},z) - v_m(X,y_{c})
$ is the error signal at a specified distance $y_{c}$ from the wall and defined by the difference between the wall-normal velocity solution $v(X,y_c,z)$ obtained from BREs and the spanwise averaged velocity $v_m(X,y_c)$; on the other hand, when wall shear stress is applied as a control variable, $e(X,z) = \tau_w(X,y_{c},z) - (\tau_w)_m(X,y_{c})$, where $(\tau_w)_m$ being the spanwise averaged wall shear stress. The amplitude and the shape of the wall deformations can be updated at each iteration based on the control signal as $
\mathscr{F}(X,z) = \mathscr{F}(X,z) + \mathscr{A}(X,z)
$. The proportional gain, $K_p$, can be determined, for example, by the frequency response method of Ziegler and Nichols \cite{Ziegler}. According to this method, the controller must be initiated with a small value of $K_p$. The proportional gain must be adjusted until a response is obtained that produces continuous oscillations; this is known as the ultimate gain, $K_{\pi}$. The desired proportional gain will then be half of the ultimate gain.

\begin{figure}
 \begin{center}
     \includegraphics[width=10cm]{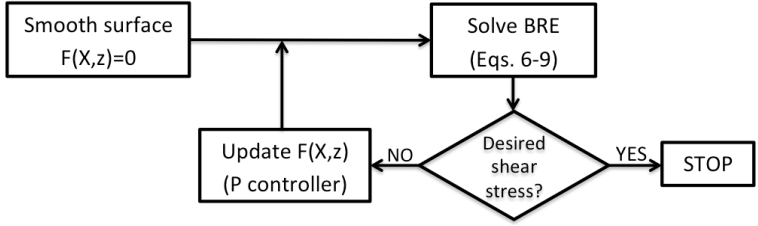}
 \end{center}
  \caption{\label{} Control algorithm schematic.}
  \label{sch}
\end{figure}

The control algorithm consists of the next steps: 
i) solve the transformed BRE numerically for the initial smooth wall surface, $\mathscr{F}(X,z)=0$;
ii) deform the wall surface using the streamwise velocity disturbance distribution on a control plane or the wall shear stress as control variable;
iii) solve the transformed BRE with the new deformed wall surface, $\mathscr{F}(X,z) \neq 0$;
iv) repeat the previous two steps until convergence is achieved (e.g., when the difference between the spatially integrated control variable from one iteration to the next is smaller than a given threshold). A schematic of the procedure is given in figure \ref{sch}.

\section{Results and discussion}\label{results}

In this section, we demonstrate the effectiveness of the control algorithm in action by applying it to G\"{o}rtler vortices excited by upstream conditions (as derived in Goldstein et al. \cite{Goldstein1}) provided by a row of roughness elements (having a Gaussian shape in $(X,z)$ plane) located at $0.5$ m from the leading edge (see figure 1 in Sescu and Thompson \cite{Sescu1}). The control variable (wall-normal velocity $v$) is taken from a sectional plane $y = y_c$ inside the boundary layer at the elevation $y_c = 0.1$ (this particular elevation was found to provide the highest reduction of vortex energy in our numerical experiments). In addition, the control based on information from the wall (wall shear stress) is considered as a comparison, since this approach is more amenable to experiments. The control is initiated in $X=0.022$ from the roughness location with a smooth ramping function.

\subsection{Numerical algorithm and grid convergence}

The numerical algorithm used to march equations (\ref{neq1})-(\ref{neq4}) in the streamwise direction is the same as the one employed in Sescu and Thompson \cite{Sescu1}. To briefly summarize, a staggered grid is used in the $y$-direction to avoid decoupling between the velocity and pressure, with second-order accurate difference schemes are employed along both $y$ and $z$ directions. The number of grid points in the wall-normal direction is $200$ and the number of grid points in the spanwise direction is $40$ (the upper boundary is at $y=20$) The marching along the streamwise $X$-direction is realized using an explicit Euler method with a sufficiently small step to avoid numerical instabilities. Since the equations are nonlinear, the convergence was achieved by a relaxation method, using an appropriate preconditioning technique applied to the first equation (\ref{neq1}). The relaxation method is based on adding pseudo-time derivatives to the original equations, and iterating until the convergence is achieved (i.e., the pseudo-time derivatives go to a very small number)

\begin{figure}
\psfrag{Lww}{$L_{ww}$}
\psfrag{si}{$\sigma$}
\psfrag{al}{$\alpha$}
 \begin{center}
   \includegraphics[width=5.5cm]{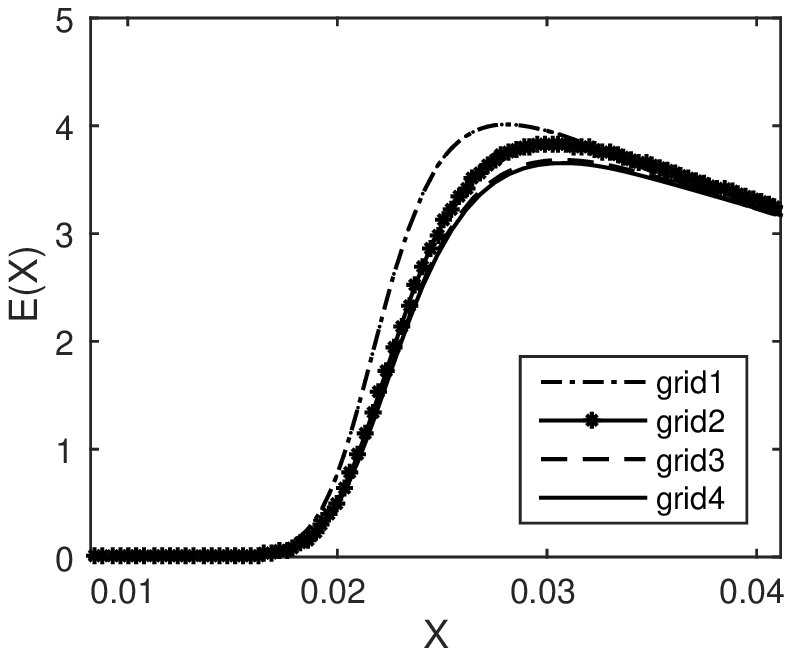} 
   \includegraphics[width=5.5cm]{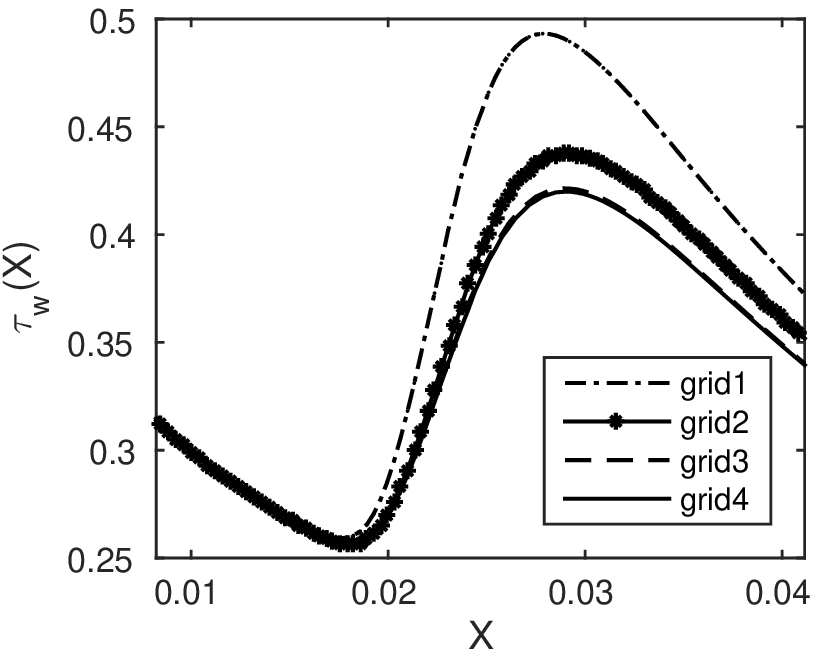} \\
a)  \hspace{55mm}  b) \\
 \end{center}
  \caption{\label{} Energy (left) and the spanwise averaged wall shear stress (right) calculated using four different grids.}
  \label{f1}
\end{figure}

In order to strengthen the level of confidence in the numerical results, a grid study is performed to determine the appropriate number of grid points along the wall-normal and spanwise directions. To this end, we consider four grid resolutions with the number of grid points given in table 1 ($N_y$ and $N_z$ correspond to the number of points in the wall-normal and spanwise directions, respectively). To reduce the computational cost, the domain size in the spanwise direction has been reduced to half distance between two roughness elements, with symmetry conditions - instead of periodic conditions - applied at the right and left boundaries. This is based on the assumption that the spanwise wavenumber associated with the vortices may be increased by nonlinear effects, while keeping the symmetry condition valid. Results in terms of energy of the disturbance and spanwise averaged shear stress distributions in the streamwise direction are plotted in figure \ref{f1} for all four grids. The solution appears to converge to the same state as the number of grid points is increased, which suggests that the resolution corresponding to 'grid3' is sufficient to obtain the desired accuracy. The numerical solution was found to be less sensitive to the resolution in the streamwise direction, for which a marching procedure is utilized to update the the flow variables.

\begin{table}[htpb]
 \begin{center}
  \caption{Number of points in different grids.}
  \label{t1}
  \begin{tabular}{rrrrrr} \hline
       Grid & $grid1$ & $grid2$ & $grid3$ & $grid4$  \\\hline
       $N_y$  &  141  & 161  & 181  & 201 \\
       $N_z$ &  21  & 31  & 41  & 51 \\
\hline
  \end{tabular}
 \end{center}
\end{table}

\subsection{Control of G\"{o}rtler vortices for two spanwise separations}

\begin{figure}
 \begin{center}
   \includegraphics[width=5.5cm]{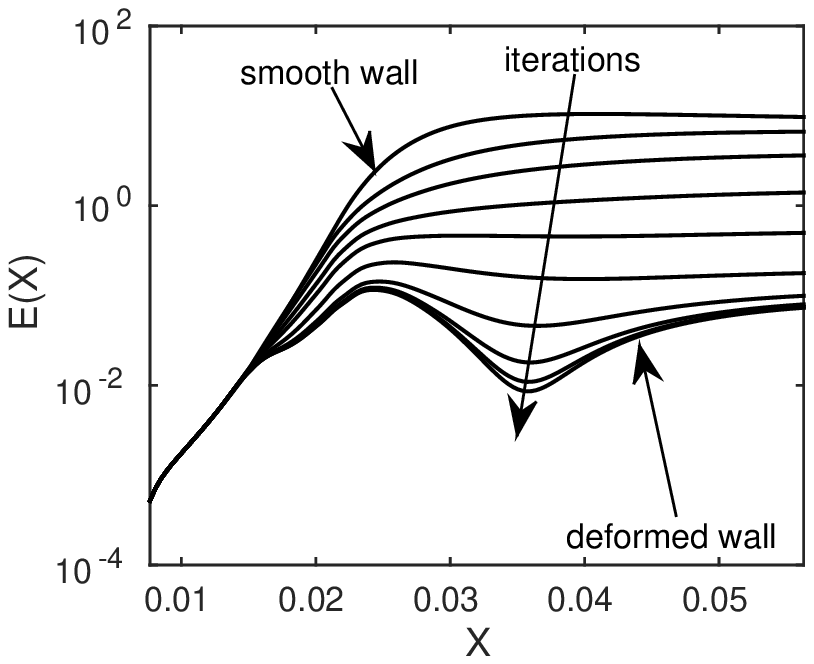} 
   \includegraphics[width=5.5cm]{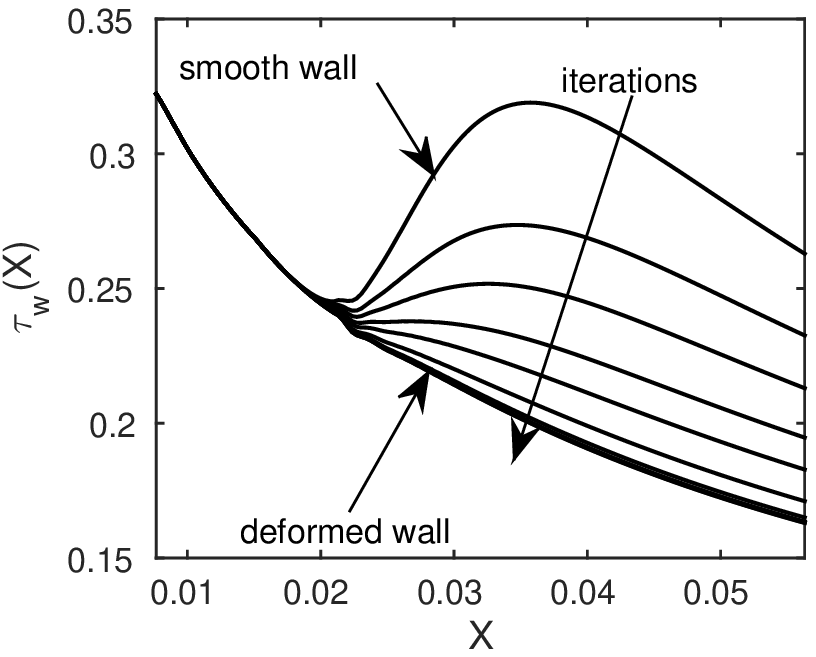} \\
a)  \hspace{55mm}  b) \\
 \end{center}
  \caption{\label{} Energy (left) and the spanwise averaged wall shear stress (right) for different iterations.}
  \label{f2}
\end{figure}

Two different spanwise separation distances of $1.2$ and $2.4$ cm, are considered, for which the G\"{o}rtler number based on the momentum displacement thickness is kept constant at $6.428$. The results are presented in nondimensional form, where the wall-normal and spanwise coordinates are scaled by the roughness spanwise separation, the streamwise coordinate is scaled by $\Lambda^{*} R_{\Lambda}$, and velocity is scaled by the freestream velocity. In the control algorithm, the number of steps necessary to reach the convergence for different cases was in the order of 10. In figure \ref{f2}, a typical convergence for both the energy of the disturbance and the wall shear stress is displayed. The energy of the disturbance was calculated from the integral

\begin{equation}\label{jj}
E(X) = \intop_{0}^{\Lambda}  \intop_{0}^{\infty}  \left[ \left| u(X,y,z) - u_m \right|^{2} +  \left| v(X,y,z) - v_m \right|^{2} +  \left| w(X,y,z) - w_m \right|^{2} \right] dzdy,
\end{equation}
where $u_m(X,y)$, $v_m(X,y)$, and $w_m(X,y)$ are the spanwise mean components of velocity.

\begin{figure}
 \begin{center}
   \includegraphics[width=12.5cm]{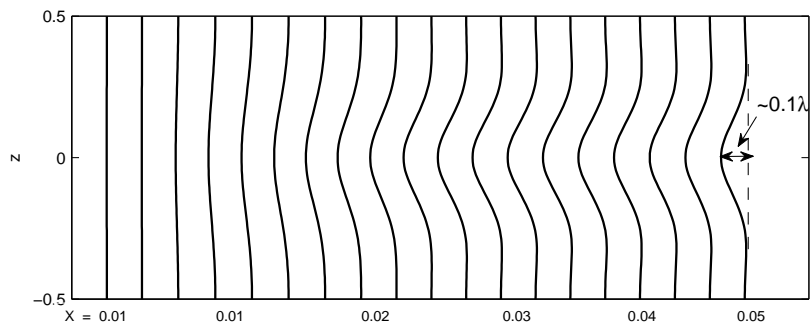}
 \end{center}
  \caption{\label{} Typical distributions of wall displacement along the spanwise direction at different streamwise locations. The maximum displacement is in the order of $0.1$ from the spanwise separation, $\Lambda^{*}$.}
  \label{f3}
\end{figure}

\begin{figure}
 \begin{center}
      \includegraphics[width=5.8cm]{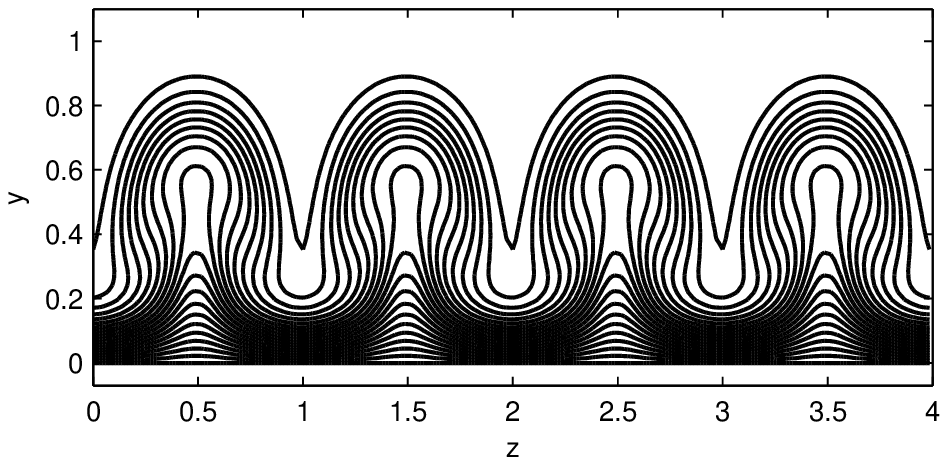}
      \includegraphics[width=5.8cm]{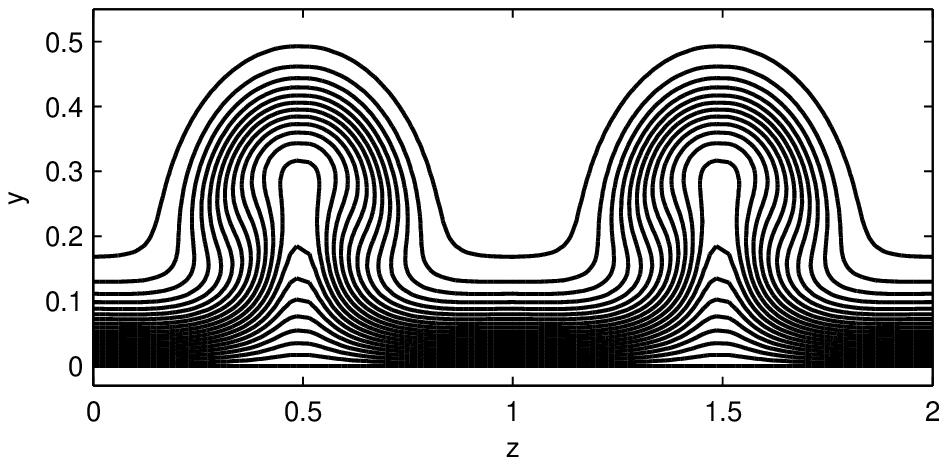}\\
a)  \hspace{60mm}  b) \\
      \includegraphics[width=5.8cm]{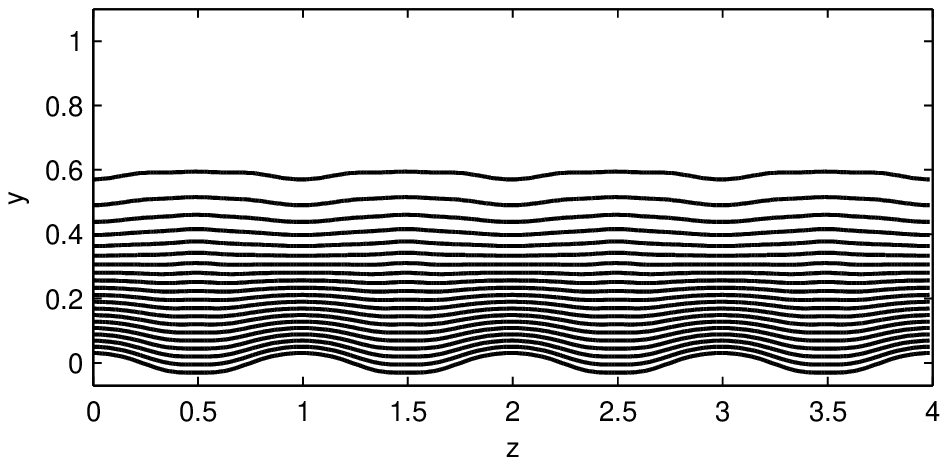}
      \includegraphics[width=5.8cm]{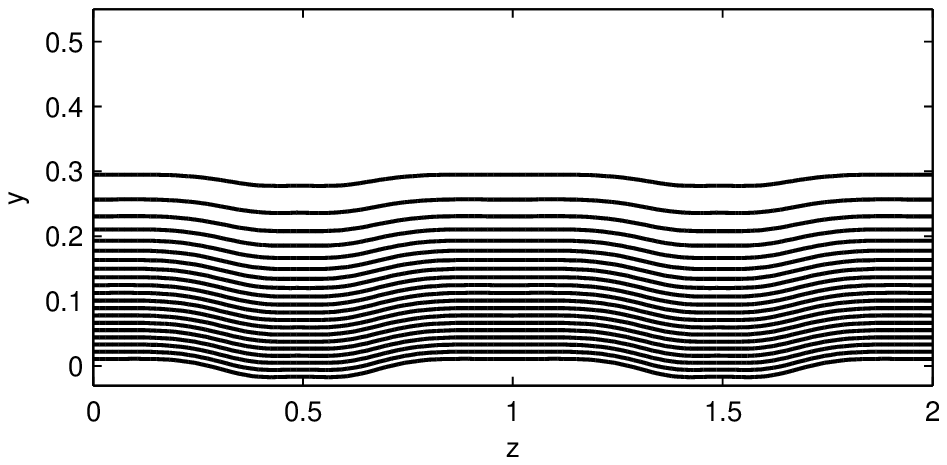}\\
c)  \hspace{60mm}   d) \\
      \includegraphics[width=5.8cm]{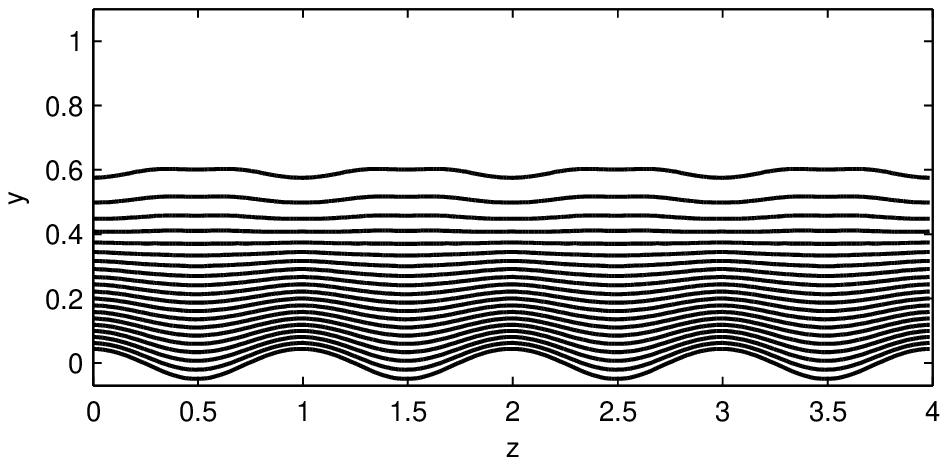}
      \includegraphics[width=5.8cm]{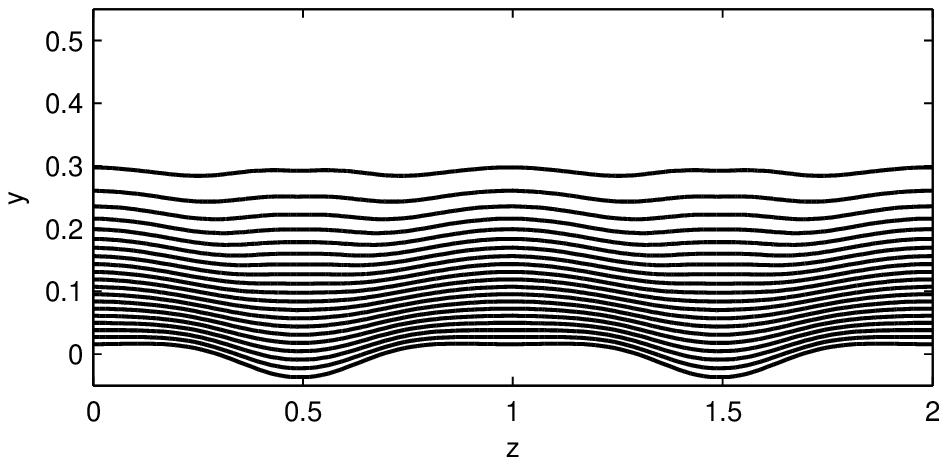}\\
e)  \hspace{60mm}   f) \\
 \end{center}
  \caption{\label{} Streamwise velocity contours for spanwise separation of $1.2$ cm (left) and $2.4$ (right): a) and b) smooth surface; c) and d) control based on $v$; e) and f) control based on wall shear stress.}
  \label{f4}
\end{figure}

In figure \ref{f3}, the deformation of the wall for control based on wall-normal velocity and $\Lambda^{*}=2.4$ cm is presented; the spanwise profiles of the displacement were plotted for different streamwise locations. It reveals that the smooth variation of the wall deformation in the streamwise direction (for different values of $X$) and that the maximum displacement of the wall is in the order of $10\%$ of the spanwise separation. Figure \ref{f4} shows contours of streamwise velocity at various cross-sections through a particular G\"{o}rtler vortex ($X=0.164$ for spanwise separation $1.2$ cm and $X=0.055$ for $2.4$ cm). The top figures \ref{f4}a and \ref{f4}b reveal G\"{o}rtler vortices developing over the undeformed surface as fully-developed `mushroom` shapes with alternating low- and high-speed streaks. Parts c and d of figure \ref{f4} correspond to results obtained from the iterative algorithm with $v$ as the control variable, while parts e and f correspond to control based on wall shear stress. The shape of the wall displacement is shown at the bottom of figures \ref{f4}c, d, e, and f. In the control algorithm, the wall surface is gradually moved downward at the spanwise location corresponding to the low-speed streak, while at the same time it is moved upward at the spanwise location corresponding to high-speed streaks. This change in geometry of the wall surface in the spanwise direction either increases or decreases the momentum of the flow, thus reducing the energy associated with the G\"{o}rtler vortices. It is obvious that both control variables (wall-normal velocity and wall shear stress) are effective in reducing the vortex strength.

\begin{figure}
 \begin{center}
      \includegraphics[width=5.8cm]{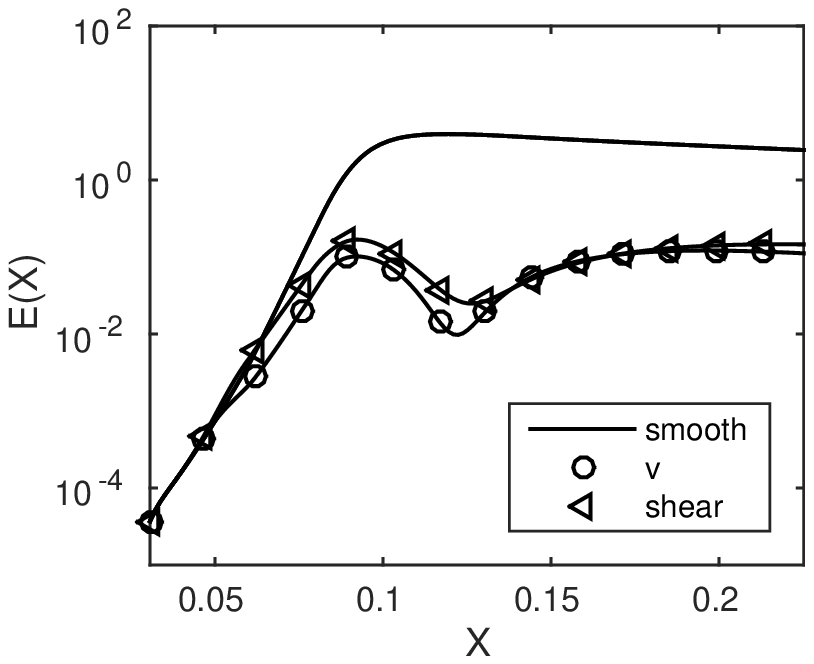}
      \includegraphics[width=5.8cm]{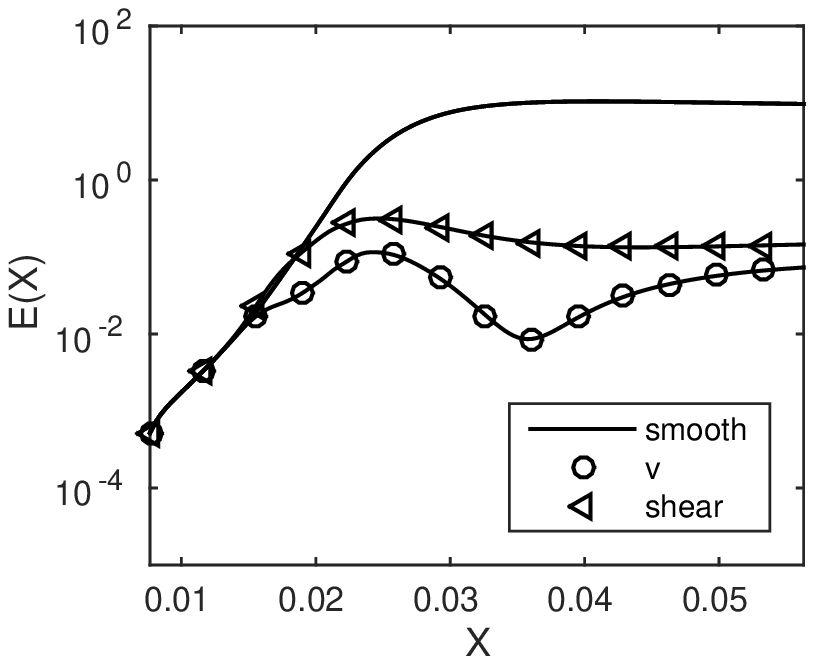}\\
a) \hspace{48mm} b) 
 \end{center}
  \caption{\label{} Energy of the disturbance as a function of streamwise coordinate: a) $\Lambda^{*} = 1.2$ cm; b) $\Lambda^{*} = 2.4$.}
  \label{f5}
\end{figure}

\begin{figure}
 \begin{center}
      \includegraphics[width=5.8cm]{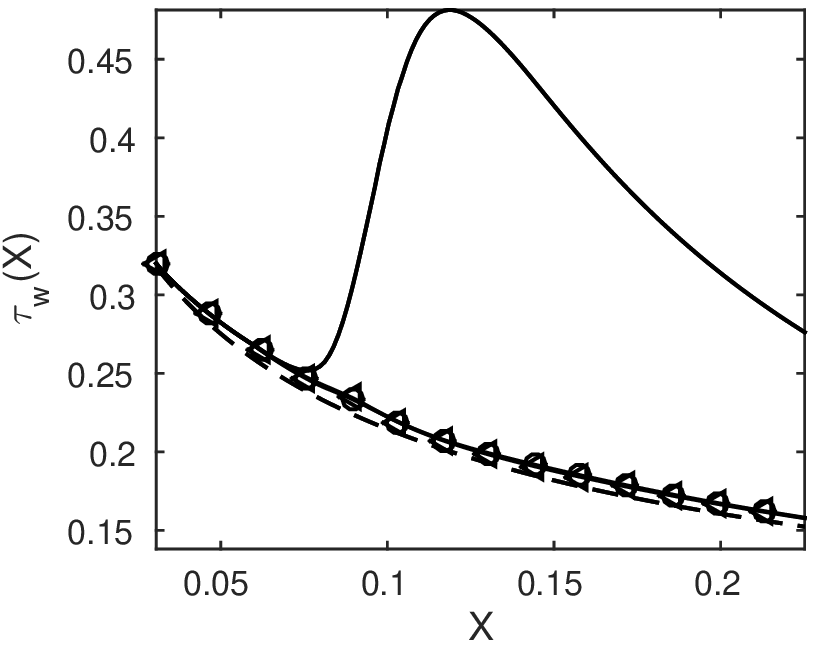}
      \includegraphics[width=5.8cm]{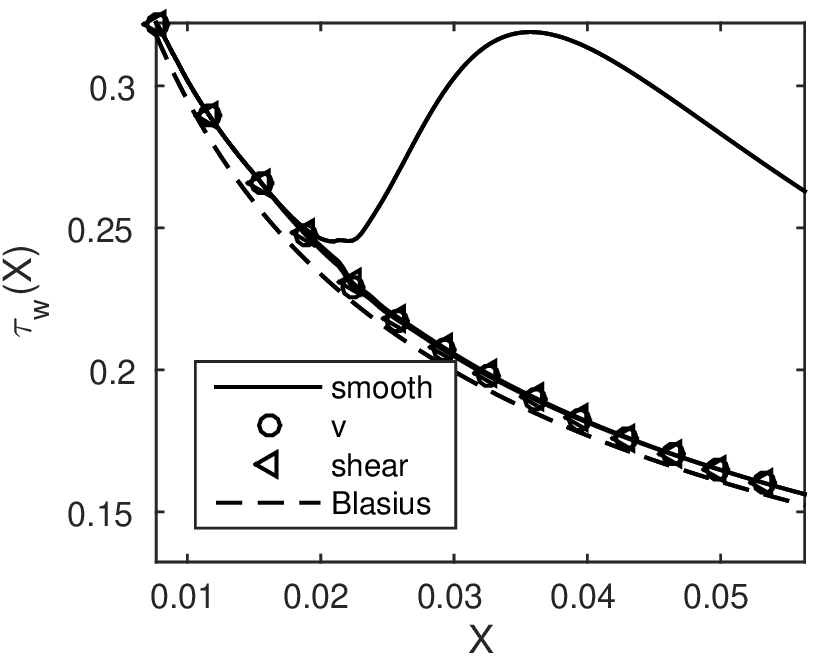}\\
a)  \hspace{60mm}  b) \\
 \end{center}
  \caption{\label{} Spanwise averaged wall shear stress as a function of the streamwise direction: a) $\Lambda^{*} = 1.2$ cm; b) $\Lambda^{*} = 2.4$ cm.}
  \label{f6}
\end{figure}

Consistent with the contour plots shown in the previous figure \ref{f5}, figure \ref{f6} shows that the kinetic energy of the disturbance is commensurately reduced. It can be observed that the energy associated with the disturbances has been significantly reduced (by almost two orders of magnitude). For the spanwise separation of $1.2$ cm (figure \ref{f5}a), the reduction in energy is almost the same for both control schemes (based on $v$ or $\tau_w$), while for the larger spanwise separation (figure \ref{f5}b) the control based on $v$ appears to be more effective. To get an idea of what effect the control has on the wall shear stress distribution in the streamwise direction (commonly used to evaluate the frictional drag), figure \ref{f6} shows the spanwise averaged wall shear stress for both the original undeformed wall and deformed wall, as well as the shear stress corresponding to the Blasius solution, corresponding to the flat plate (i.e., no surface curvature). Both subfigures show that the shear stress at the wall is considerably reduced and approaches the Blasius solution, which is an indication that the frictional drag can be significantly affected using this control scheme.

\begin{figure}
 \begin{center}
   \includegraphics[width=5.5cm]{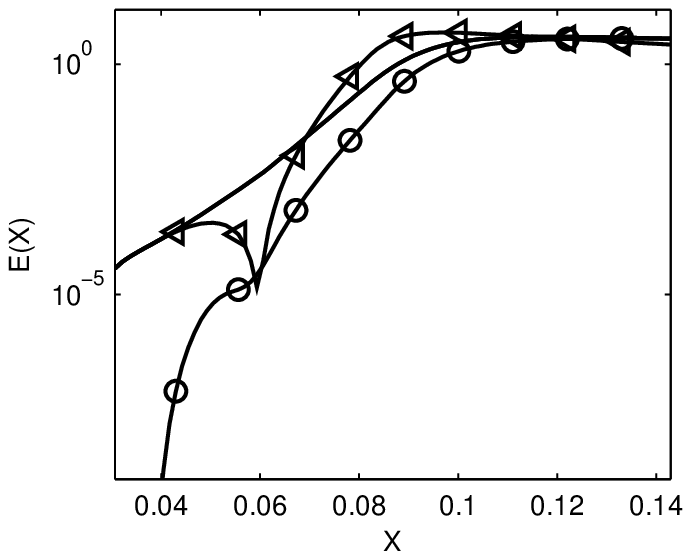} 
   \includegraphics[width=5.5cm]{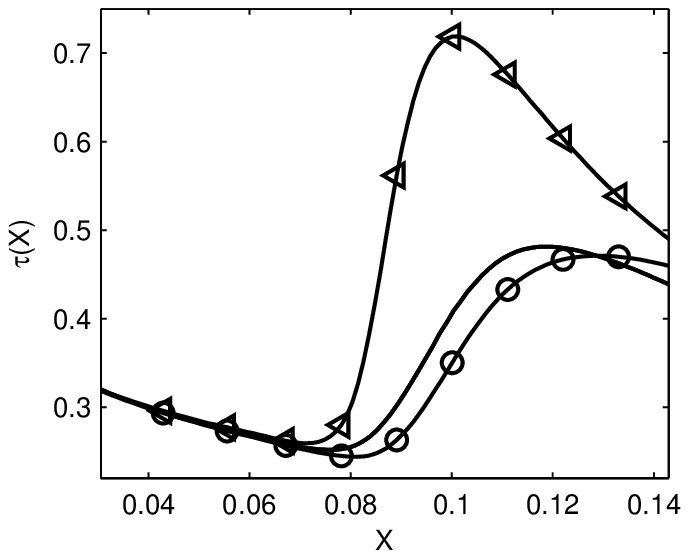} \\
a)  \hspace{55mm}  b) \\
 \end{center}
  \caption{\label{} Energy (left) and the spanwise averaged wall shear stress (right): solid lines - the smooth surface with upstream conditions; circles - deformed wall without upstream conditions; triangles - with upstream conditions, and the deformed wall translated $\Lambda^{*}/2$ along the spanwise direction with respect to the vortex axis.}
  \label{f7}
\end{figure}

Once the control algorithm has reached a converged solution, it is interesting to find what the effect of wall deformations is in the absence of any upstream conditions. This amounts to de-activating, so to speak, the upstream conditions that model the flow near the roughness elements, whilst allowing wall deformations to persist in the downstream, and marching the BRE along the streamwise direction. Figure \ref{f7}a shows a comparison between the energy associated with the aforementioned calculation (circle symbols) and the original energy corresponding to the smooth surface and vortices excited by Goldstein et al. \cite{Goldstein1} upstream conditions (solid line). It appears that G\"{o}rtler vortices generated by wall deformations alone would carry the same amount of energy in the saturation region, although the energy increase in the region upstream of the saturation point is different for each case. The wall shear stress plot in (\ref{f7}b) shows that wall deformations provide reduced shear stress up to $X=0.127$ and slightly increased shear stress after that point.

Another interesting experimental calculation that can be performed after the control algorithm has reached the final solution involves a shift of the wall displacement in the spanwise direction by $\Lambda^{*}/2$. This amounts to marching the BRE with upstream conditions from Goldstein et al. \cite{Goldstein1}, where the deformed surface wall (from the last iteration) is translated along the spanwise direction by half of the spanwise separation. Figure \ref{f7} includes the energy and wall shear stress distributions from such a calculation (triangle symbols). As expected, there is a higher amplification of energy compared to the smooth surface because in this case the momentum is injected into the low-speed streak, and absorbed from the high-speed streak (opposite to the controlled case). This produces a significant increase in the wall shear stress compared to the smooth surface, as shown in figure \ref{f7}b (triangle symbols versus solid line).

\vspace{2mm}

\section{Conclusions}

A numerical study of the effect of controlled wall deformations on G\"{o}rtler vortices development for an incompressible boundary layer flow over a curved wall has been performed. The problem was formulated in a high Reynolds number asymptotic framework, in which the streamwise vortex flow is determined by the BRE (\ref{nq1})-(\ref{nq4}) together with appropriate upstream boundary conditions. The effect of wall deformations was incorporated into the BRE's through a Prandtl transformation of dependent and independent variables in the wall normal direction; the BRE's were then solved numerically using a marching algorithm. The transformed BREs eqns (\ref{neq1})-(\ref{neq4}) then explicitly included an arbitrary function $\mathscr{F}(X,z)$ representing the local surface deformation. The proposed simple control strategy, aimed at minimizing the vortex energy for a given $\mathscr{F}(X, z)$, used either the wall-normal velocity disturbance at a specified height from the wall or the wall shear stress as the control variable.

A simple test case involving a boundary layer developing over a concave surface that is excited by a row of roughness elements was considered here (the algorithm can be extended to other boundary layer flow scenarios, including flat-plates with roughness elements or/and freestream disturbances). Our results clearly indicate that the controlled surface deformations are rather effective in altering the streamwise development of the G\"{o}rtler vortices for a given spanwise wavelength. The kinetic energy associated with the vortices was shown to decrease by almost two orders of magnitude from the original amplitude. The spanwise averaged wall shear stress distribution in the streamwise direction (commonly used to evaluate the frictional drag) was plotted for both the original undeformed and deformed wall, as well as corresponding solely to the Blasius solution. Results using either control variables showed that the shear stress at the wall can be reduced considerably, approaching the Blasius solution. This is an indication that the frictional drag can be significantly reduced.

\section*{Acknowledgments}

The authors gratefully acknowledge comments and suggestions from Dr. David Thompson from Mississippi State University. We would also like to acknowledge partial support from NASA EPSCoR RID Program through Mississippi State Grant Consortium directed by Dr. Nathan Murray. We would like to thank the anonymous reviewers for very constructive comments and suggestions. MZA would like to thank Strathclyde University for financial support from the Chancellor's Fellowship.


\end{document}